\documentclass[sigconf]{acmart}
\AtBeginDocument{
  \providecommand\BibTeX{{
    \normalfont B\kern-0.5em{\scshape i\kern-0.25em b}\kern-0.8em\TeX}}}

\setcopyright{none}
\copyrightyear{2023}
\acmYear{2023}
\acmDOI{XXXXXXX.XXXXXXX}
\acmConference[HRI '23]{HRI 2023}{March 13--16, 2023}{Stockholm, Sweden}
\acmPrice{15.00}
\acmISBN{978-1-4503-XXXX-X/18/06}

\usepackage{dramatist}

\begin{document}

\title{Utilising Explanations to Mitigate Robot Conversational Failures}

\author{Dimosthenis Kontogiorgos}
\email{kontogiorgos@uni-potsdam.de}
\affiliation{
  \institution{University of Potsdam}
  \country{Germany}
}

\renewcommand{\shortauthors}{Dimosthenis Kontogiorgos}

\begin{abstract}
This paper presents an overview of robot failure detection work from HRI and adjacent fields using failures as an opportunity to examine robot explanation behaviours. As humanoid robots remain experimental tools in the early 2020s, interactions with robots are situated overwhelmingly in controlled environments, typically studying various interactional phenomena. Such interactions suffer from real-world and large-scale experimentation and tend to ignore the \emph{`imperfectness'} of the everyday user. Robot explanations can be used to approach and mitigate failures, by expressing robot legibility and incapability, and within the perspective of common-ground. In this paper, I discuss how failures present opportunities for explanations in interactive conversational robots and what the potentials are for the intersection of HRI and explainability research.
\end{abstract}


\begin{teaserfigure}
\includegraphics[width=\textwidth]{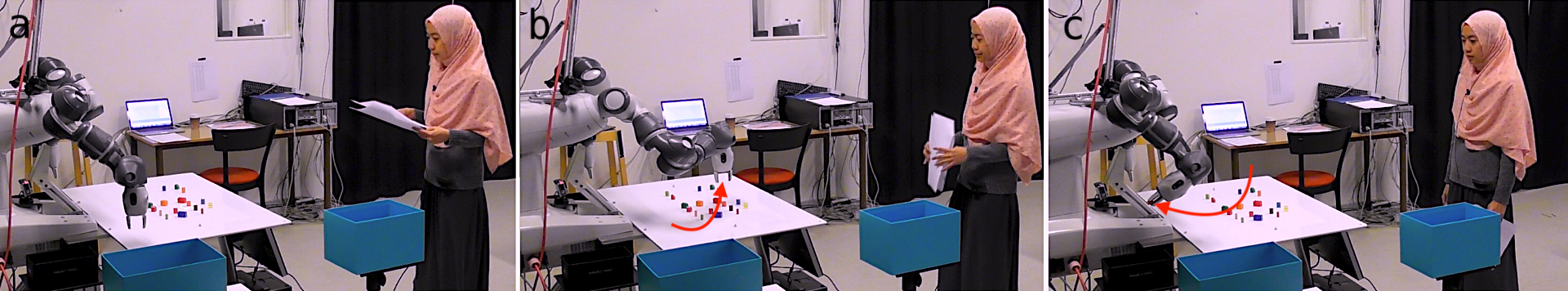}
\caption{A human user instructing a robot dual-arm to pick-and-place objects: a) the human utters an instruction, b) the robot attempts to grasp the object, c) the robot indicates incapability through sudden arm movement. Even though the robot does not have a head and cannot speak, it affords interactional phenomena through non-verbal behaviour. Experiment published at \cite{sibirtseva2018comparison}.}
\Description{A human and a robot interacting in a pick-and-place task.}
\label{fig:yumi}
\end{teaserfigure}

\maketitle

\section{Introduction}
\begin{figure*}[t!]
\centering
\includegraphics[width=\textwidth,scale=1]{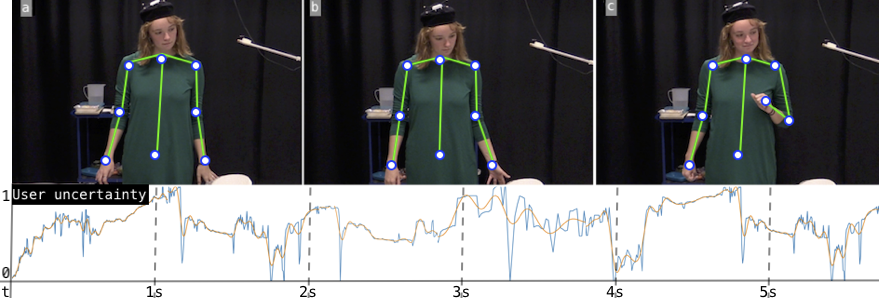}
\caption{User uncertainty estimation in response to a robot failure demonstrated in non-verbal signals \cite{kontogiorgos2019estimating}. Deviating from expected behaviour, the user is unable to follow the robot's goal, eliciting the need for explanation \cite{kontogiorgos2021systematic}. The user here does not ask for clarification but her embodied actions indicate signs of misunderstanding, including a smile and gaze towards the robot.}
\label{fig:action}
\end{figure*}

\emph{Filippos} is a goldsmith, \emph{Jakob} is his robot assistant. They work together:

\begin{drama}
  \Character{Filippos}{filippos}
  \Character{Jakob}{jakob}

  \filipposspeaks: \direct{raises eye-brows, looks at the instructions} All right computer, let's get it right this time!
  \jakobspeaks: \direct{frowns} Don't call me that!
  \filipposspeaks: \direct{winks} Sorry!
  \jakobspeaks: \direct{smiles} Okay, what are we making today?
  \filipposspeaks: \direct{looks again at instructions} This client wants us to create a custom-made necklace, \emph{it should not look too shiny, and it should not look too `cheap' either}.
  \jakobspeaks: \direct{natural-language processing unit produces low confidence on this definition, \textbf{\textcolor{red}{rolls eyes, makes beep-bop sound}}}
  \filipposspeaks: \direct{Filippos notices Jakob's confusion}
  \jakobspeaks: \direct{looks at Filippos} Filippos, what's \emph{`not too cheap'}?
  \filipposspeaks: \direct{raises eye-brows} I don't know either, but let's try this for a minute. Can you hold the Vernier caliper, please?
  \jakobspeaks: \direct{computer-vision unit detects caliper in a position too close to grasp, \textbf{\textcolor{red}{Jakob moves its arm twice indicating incapability to grasp item}}} \textbf{\textcolor{red}{Oh-oh!}}
  \filipposspeaks: \direct{looks at Jakob, passes on the caliper} Oh! My bad, it is too close to you, there you go.
\end{drama}

\vspace*{-5mm}
\noindent
While such a rich interactional setting with a robot seems out of reach in early 2023, one can imagine such mechanisms will be expected as machines acquire language skills. Certainly, there are interaction expectations that need to be fulfilled too, once robots afford such conversational phenomena. What this story was designed to illustrate however, is the robot's behaviour and, in particular, behavioural elements expressing robot incapability or making its ability to understand more transparent. The robot here encounters either sensory or computational failures but it is able to explain in human terms what has gone wrong. Explanations in this view, are not only justifications for its actions but also embodied demonstrations of mitigating failures by acting through multimodal behaviours (in the text above marked in \textbf{\textcolor{red}{red}}).

Explaining the reasons for failures significantly affect robots' ability to establish mutual understanding with users \cite{kontogiorgos2022mutual}. Failures and explanations should be examined from the perspective of common ground; robots should generate explanations utilising language along with multimodal behaviours, making more transparent their state of understanding. There is a lack of explainability work tailored to mitigating robot failures, especially with users not knowledgeable of how these systems work. Overall, existing approaches conceive explanations as statistical properties decoupled from dynamic user environments without knowing whether an explanation is in fact needed, or how to best convey it to users. In this interactive approach, explanations are placed at the same level as other communicative acts, complementary to the interpretability perspective of making statistical causality more transparent.

\section{Detecting Robot Failures}
\begin{figure*}[t!]
\centering
\includegraphics[width=\textwidth,scale=1]{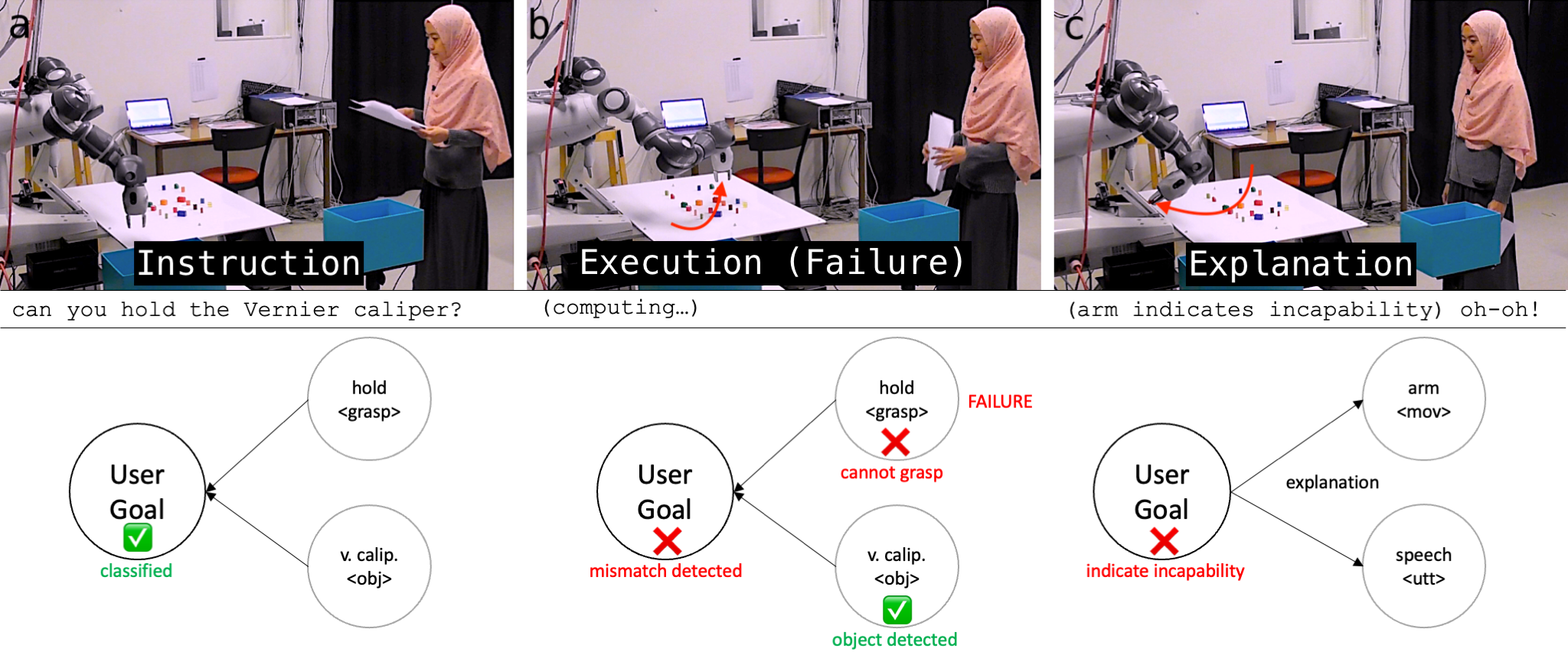}
\caption{A robot utilising non-verbal behaviour to indicate task incapability after a failure has been detected to successfully respond to the user's intent. The robot is making legible its intention to comply but also its inability to do so due to a failure.}
\label{fig:explanation}
\end{figure*}

\emph{`Many of the errors that occur in human–computer interaction can be explained as failures of grounding'} (Brennan \cite{brennan1998grounding}). While a lot of HRI work has focused on what are the effects on the human perception of the robot with regards to failure \cite{honig2018understanding}, less research has been conducted on how to automatically detect the failure and the reasons for human-robot misalignment in communication \cite{kontogiorgos2021systematic}.

Failures in human-robot communication can be interpreted as \emph{deviations from expected behaviour}. From the user's perspective, robot failures often violate social protocols of interaction, such as not responding or failing to comply with user requests (Figure \ref{fig:yumi}). An important distinction in HRI is that multimodality is fundamental for failure detection as uncertainty in user behaviour may not always be explicitly verbalised \cite{kontogiorgos2019estimating} (Figure \ref{fig:action}). Human reactions to robot failures seem to vary but they are nevertheless predictable \cite{honig2018understanding}. Signal variations exist in verbal and non-verbal cues, such as \emph{eye-gaze} and \emph{head movement}, \emph{facial expressions}, \emph{body motion}, as well as \emph{speech} and \emph{acoustic} features \cite{kontogiorgos2022mutual}. Deviations in user behaviour can also be modelled as a lack of social contingency, through low-level sensor-input features \cite{short2018detecting}.

The open challenges in detecting robot failures are consequently twofold: \emph{first}, the robot needs to detect a failure has occurred, and \emph{second}, it needs to be able to recover from the (detected) failure, thereby conveying social intelligence (a very challenging human-like behaviour). Robot failures also have an impact on the development of user trust \cite{porcheron2022definition}, a highly influential dimension that is also regulated by embodiment and system performance. Trustworthiness is highly affected by failure mitigation strategies and how the robot utilises explanations to justify the reasons for a failure \cite{kontogiorgos2022mutual}.

Some HCI and HRI work has focused on the development of \emph{frameworks} \cite{wang2019designing,rohlfing2020explanation,sanneman2022situation,milani2022survey} for how human explanations can be applied in XAI research, as well as on \emph{empirical observations} of how robots should explain and mitigate failures to users \cite{correia2018exploring,kontogiorgos2020behavioural,thielstrom2020generating,van2022correct}. Robot failures, in particular, present an essential exploratory process of how to provide contingent explanations, especially when robots attempt to inform users on why they are unable to accomplish requested tasks \cite{kwon2018expressing,han2021need} (Figure \ref{fig:explanation}).

\section{How is XAI Relevant?}
The ability to explain (\emph{explainability}) forms a significant factor to the development of \emph{trust} in artificially intelligent systems \cite{kim2015ibcm,ribeiro2016should,yang2017evaluating}, as it conveys understanding of the system's own actions and further develops users' perception of reliability in the system. From an \emph{interactive} point of view, when users ask a robot to justify its actions (and thereby its failures), it should be able to respond with an intelligible explanation \cite{bryson2017standardizing}, satisfying not only \emph{algorithmic transparency} but also conform to \emph{social protocols of interaction}. This leads to the need for robot \emph{explanation interfaces} that are able to determine how to best mitigate failures to the current user and in what format such explanations should be.

Two main branches of XAI research are the \emph{interpretability} of ML models (in terms of their transparency) and the \emph{justification} of their prediction (answers to \emph{`why'} questions) \cite{biran2017explanation,gilpin2018explaining}. In this paper, I discuss the transparency dimension as a proxy for mutual understanding in HRI, and less the justification dimension that has impact on how to give reasons for decisions taken. In particular, I emphasise the importance of why robots should automatically generate explanations utilising natural-language along with classification predictions utilising their sensory input (i.e., Figure \ref{fig:explanation}).

I highlight the notion of \emph{transparency} in particular, as explanations may differ in nature depending on who is asking for an explanation. A medical-robot developer may have a different need for explanation on the robot's failure on a diagnosis than a patient or a doctor would \cite{bryson2017standardizing}. That means that different levels of explanation abstraction need to be presented according to who the robot is interacting with \cite{langley2017explainable}. This is similar to how humans estimate each other's knowledge to align the information uttered and establish common ground: whether you are talking to (a) your extended family, (b) your friends, or (c) your colleagues, you are probably working with (a) \emph{`computers'}, (b) \emph{`human-computer interaction'}, or (c) \emph{`human-robot interaction research with an interest in the linguistic aspects of human-robot communication'} \cite{fussell1992coordination}. Explanations in this view are co-constructed through an adaptation process that socially intelligent speakers can easily adjust.

Understanding adaptation from the HRI perspective means that we can imagine situations in which utilising explanations to mitigate a robot failure may take several forms, and the verbal channel may not always be the most appropriate channel. Post-hoc explanations may be formed through embodied demonstrations, whether it is through movement, gestures, or eye-gaze, indirectly pointing to the reason for failure. Generating explanations in this form implies that the robot may need to attend to the user state to determine whether an explanation for its action is needed without always waiting for an explanation prompt (i.e., \emph{`why did you do that'}). From the analytical point of view, XAI techniques (i.e., post-hoc explanations) also have large implications on highlighting the markers that the autonomous robot needs to pay attention to in order to detect, classify, and resolve failures \cite{weber2020towards}, as well as give insights for why did a robot take certain decisions.

\section{Explanations as Communicative Acts}
The interactive approach of explainability considers explanations to be communicative acts, which differs from the interpretability perspective of making statistical causality more transparent. The social nature of modern technological interfaces makes the need for explanations through \emph{natural-language} essential \cite{rohlfing2020explanation}, as users will expect to receive explanations similarly to how they would receive explanations from humans. In fact, utilising natural-language as the principal medium of interaction introduces the problem of \emph{mutual understanding} at the centre of HRI failure and miscommunication explanation behaviours \cite{kontogiorgos2022mutual}. Existing approaches outside the field of HRI conceive explanations as \emph{autonomous processes} decoupled from \emph{dynamic user environments}, neither knowing whether an explanation is needed nor how to best convey it.

There is currently a lack of data-driven methods in: a) \emph{how to detect in real-time the need for explanations}, b) \emph{generate explanations visually grounded to the user's environment} and \emph{adapted to the user's information needs}. In situated human-robot interactions, utterance production is a \emph{highly collaborative and participatory process}; robot failure explanations should as well be adapted and formulated to the user's information needs and concurrent to the changes in the \emph{shared space of attention}. Such adaptation strategies will allow humans to act collaboratively and more efficiently (i.e., required amount of turns spoken) than in non-interactive settings.

This paper proposes the investigation of incremental production strategies of explanations in HRI. I take in this context the \emph{traditional} view of explainability, where robot/algorithmic transparency can be used as a medium to assist and navigate the grounding process. It can also be used as a tool to communicate the degree of a robot's uncertainty, making robots' intent more transparent \cite{kwon2018expressing}. In this view, statements needed to clarify robot legibility or incapability manifest that the utterances spoken or the non-verbal signals expressed adjust any differences between the user and the robot that may cause failures or misunderstandings. This approach does not involve explainability in the form of justifications \cite{scheutz2022transparency}, yet it does involve the indication of \emph{reasons} for whether something is understood or misunderstood. It also involves the ability of the agent to \emph{explain the causes of misunderstanding, mitigate the reasons for failures in a collaborative manner, and communicate its understanding of the user's intent and goals}.

Providing explanations in the form of probabilities or statistical relationships is probably not as effective or satisfying for users as referring to the causes for failures \cite{miller2019explanation}. Explanations as discourse units reveal intentions and can facilitate learning \cite{miller2019explanation}, and users can derive better mental models of robot behaviour when it provides causes of incapability or explain what it does not understand. Such explanation adaptation mechanisms should also follow the principles of cooperative communication \cite{grice1975logic}, putting not only the explanation properties in focus but also how it is conveyed and according to the user's degree of understanding \cite{rohlfing2020explanation}. In interactive turn-taking, where each turn is a sequential classification, a classification result that leads to a failure is never completed but something that can be continuously revised and reformed. Explanations in this view become an affordance, an interaction property of the system, that invites users to participate in co-constructing explanations and form a shared understanding of the reasons a robotic system may encounter communication failures.

\section{Future Research}
Once explanations follow such criteria, they also need to represent socially contingent actions to moments of miscommunication generated at the right place the right time (i.e., principles of grounding and turn-taking \cite{clark1991grounding}). The grounding principle here is essential because explanations require that the agent is rational and has \emph{common sense} or a common understanding of the world. Addressing these questions in situated multimodal human-robot interactions, there are open challenges in: \emph{a) identifying the key multimodal indicators on whether (and when) explanations for failures are needed, and b) investigating how such explanations should be produced collaboratively as discourse units, and c) co-constructing explanations following social protocols of human communication} \cite{miller2019explanation}.


\balance
\bibliographystyle{ACM-Reference-Format}
\bibliography{bibliography}

\end{document}